\RequirePackage[2020-02-02]{latexrelease}
\documentclass[nofootinbib]{revtex4}

\usepackage{amsmath}
\usepackage{amssymb}

\usepackage{textgreek}

\usepackage{comment}

\begin{document}

\title{\Large
On BRST-Related Symmetries in the FLPR Model with Gribov Ambiguities
}

\author{Bhabani Prasad Mandal\footnote {e-mail address: bhabani@bhu.ac.in}}

\affiliation{Department of Physics,
Banaras Hindu University,\\ Varanasi - 221005, India.}
 
\author{Sumit Kumar Rai\footnote {e-mail address: sumitssc@gmail.com}}

\affiliation{Sardar Vallabhbhai Patel College,\\
Veer Kunwar Singh University,\\Ara, Bhabua - 821101, India.}

\author{Ronaldo Thibes\footnote{e-mail address: thibes@uesb.edu.br}}
\affiliation{Departamento de Ci\^encias Exatas e Naturais,\\
Universidade Estadual do Sudoeste da Bahia,
Itapetinga - 45700000, Brazil.}

\begin{abstract}
With a recent revival, novel features of the FLPR model have been reported in the literature.  A connection between those features to QCD involving the Gribov problem is explored.
We investigate the FLPR model in a recently proposed framework of BRST-related symmetries and perform its full functional quantization as a gauge invariant system taking into account the Gribov ambiguities.  We obtain a family of BRST-related transformations generated by the discrete group of symmetries of the action.  We show that gauges possessing Gribov ambiguities lead to a violation of the initial discrete group of symmetries of the gauge-fixed action.  The obtained results shed light into similar issues in QCD by the corresponding association of variables and fields between the two systems.

\end{abstract}

\maketitle

\section{Introduction}
It is not from today that symmetries in theoretical physics are deeply intertwined in the core of its fundamental principles, leading to measurable consequences on the experimental counterpart.   To mention two emblematic examples from the micro and macro universes, both the Standard Model of particle physics and the cosmological  \textLambda CDM Model, as well as their extensions and generalizations, are firmly rooted on external and internal symmetries.   In that respect, concerning the quantum aspects of our world, one of the most prominent characteristics in its modern understanding is certainly the BRST and BRST-related symmetries \cite{Mandal:2023pdk}.  Surviving even after gauge-fixing, the original BRST symmetry 
\cite{Becchi:1974xu, Becchi:1974md, Tyutin:1975qk, Becchi:1975nq}
has played a decisive role in the proof of unitarity and renormalizability of Yang-Mills theories and is present in the quantum description of all gauge-invariant systems.  After the appearence of its anti-BRST \cite{Curci:1976bt, Curci:1976ar} and (anti-)co-BRST
\cite{Lavelle:1993xf, Tang:1994ru, Malik:1997ge}
forms, alongside with its finite-field-dependent (FFBRST) versions \cite{Joglekar:1994tq,Rai_mandal_ham, off_shell_jmp, Joglekar:1998dw,GZ_ffbrst},
BRST-related transformations represent nowadays an essential tool for quantization schemes, either from operatorial canonical or functional integral approaches.

On the other hand, the confinement of quarks and gluons in QCD configures one of the most intricate still open problems in high-energy physics.   Being an intrinsic gauge theory, it is necessary to choose a specific gauge-fixing to perform calculations in QCD.  In the seminal paper \cite{Gribov:1977wm}, V. N. Gribov pointed out serious issues related to the accessibility and unicity of gauge-fixing conditions in QCD in what came to be known as the Gribov ambiguities problem -- possibly related to the quarks and gluons confinement understanding.   That opened an interesting investigation route in which modifications in the functional integration measure have been proposed to avoid Gribov gauge copies in the quantum theory, leading to the horizon function
\cite{Zwanziger:1989mf, Capri:2006cz}
and BRST soft-breaking \cite{Dudal:2008sp, Sorella:2009vt} concepts in QCD.  Not just the Landau, but other gauges as well, were shown to suffer from Gribov ambiguities, of which we may mention the maximal Abelian gauge (MAG) \cite{Dudal:2006ib, Capri:2008ak, Deguchi:2016qge} and interpolating gauge-fixing attempts \cite{Dudal:2005zr, Capri:2006bj}. An attempt was made to connect Gribov-Zwanziger theory which is free from Gribov copies to Yang-Mills  theory which suffers with Gribov ambiguities \cite{GZ_ffbrst} through FFBRST transformation.

Connecting the two previous subjects, a recent paper by R Malik \cite{Malik:2023rze} has presented and discussed a series of new BRST-related symmetries in the Friedberg-Lee-Pang-Ren (FLPR) model, namely (co-)(anti-)BRST symmetries as well as other algebraic generated ferminonic and bosonic ones, promoting a revival of the topic in the literature \cite{ Krishna:2023plc, Nair:2023jvg, Nair:2023qfx,  Chauhan:2024nnm, Nair:2025pjt}.  The use of the FLPR system as a toy model to investigate Gribov ambiguities in a simpler framework has been proposed in \cite{Friedberg:1995ty} and subsequently further analysed in references \cite{Fujikawa:1995fb, Villanueva:1999st, Canfora:2014xca}.  With a renewed interest, after Malik showed explicit realizations of (co-)(anti-)BRST symmetries \cite{Malik:2023rze}, the FLPR model has been revisited from other perpectives and viewpoints in \cite{Nair:2023jvg, Krishna:2023plc, Nair:2023qfx, Chauhan:2024nnm, Nair:2025pjt}.  In particular, it has been shown in \cite{Krishna:2023plc, Nair:2023jvg} that the FLPR model can be connected to Hodge theory in terms of its BRST-related symmetry generators.   Modern complementary techniques have been used to approach the FLPR issues such as supervariables and supersymmetry \cite{Nair:2023jvg}, (anti-)chiral supervariable \cite{Chauhan:2024nnm}, symplectic Faddeev-Jackiw \cite{Nair:2023qfx} and BFV quantization \cite{Nair:2025pjt}.  However, in spite of exposing new features of the FLPR model, few has been added concerning its relation to Gribov ambiguities -- that is precisely one of the points we shall address here.  Our contribution ammounts to showing how the FLPR model can be cast into the recent framework of BRST-related symmetries presented in \cite{Mandal:2023pdk} and how different gauge-fixing choices affect the discrete symmetries of the action.  In particular, we shall see that Gribov ambiguity gauge-fixings violate the larger discrete symmetries group algebra \cite{Mandal:2023pdk}  restricting the possibilites for the BRST-related transformations.

Our work is organized as follows.  After this initial Introduction, in Section 2, we briefly review the characterization of the FLPR model and  introduce our notation and conventions.  In Section 3, we proceed to the functional quantization of the model {\it a la} Batalin-Fradkin-Vilkovisky, introducing the necessary ghost fields in phase space corresponding to the first-class constraints present in the system.  In Section 4, we discuss the possibilities for gauge-fixing the model and generate a family of BRST-related symmetries.  The Gribov ambiguities are discusses in Section 5, in which a corresponding restriction to the path integral is performed.  We close in Section 6 with some concluding remarks.

\section{The FLPR Model}
The idea of the FLPR model has appeared in the context of proper gauge orbits in canonical quantization in \cite{Kuchar:1986ji} and has been proposed to constitute a tool to study Gribov ambiguities by comparison to QCD in algebraic gauge-fixing scenarios in \cite{Friedberg:1995ty}.  In its mechanical version, it can be described by the Lagrangian function 
\begin{equation}\label{L}
L = \frac{1}{2}\left[
(\dot{x}+\alpha qy)^2+(\dot{y}-\alpha qx)^2+(\dot{z}-q)^2
\right]
-V(\rho)
\end{equation}
with
\begin{equation}\label{rho}
\rho\equiv \sqrt{x^2+y^2}
\,,
\end{equation}
corresponding to a well engendered system characterized by the four one-variable functions $x,y,z,q$.  Upon quantization, the FLPR model (\ref{L}) leads to a $(0+1)$ quantum field theory which can be compared to QCD with respect to its gauge-fixing properties, specially concerning the Gribov ambiguities issue \cite{Friedberg:1995ty, Fujikawa:1995fb, Villanueva:1999st, Canfora:2014xca}.   In our notation convention, in the Lagrangian function (\ref{L}), $\alpha$ denotes a convenient coupling constant, $V(\rho)$ a generic potential depending on the symmetric combination (\ref{rho}) and the dot stands for ordinary derivative with respect to the time evolution parameter.

Given an arbitrary  time-dependent function $\xi (t)$, the Lagrangian function (\ref{L}) is left invariant under the transformation
\begin{equation}\label{GI}
\begin{split}
x &\longrightarrow  x \cos (\alpha\xi)- y \sin (\alpha\xi) \,, \\
y &\longrightarrow x\sin (\alpha\xi) + y \cos (\alpha\xi) \,, \\
z &\longrightarrow z+ \xi \,, \\
q &\longrightarrow q+ \dot{\xi} \,, \\
\end{split}
\end{equation}
signaling the presence of a local symmetry.  In fact, introducing the canonical momenta $p, p_x, p_y, p_z$, respectively conjugated to $q, x, y, z$,  the canonical Hamiltonian associated to (\ref{L}) can be written as
\begin{equation} \label{Hc}
H_c= \frac{1}{2}\left( p_x^2+p_y^2+p_z^2 \right)+q\left[\alpha\left(xp_y-yp_x\right)+p_z\right]+V\left(\rho\right)\,,
\end{equation}
and, by following the Dirac-Bergmann approach \cite{Dirac:1950pj, Anderson:1951ta, Rothe:2010dzf}, the system dynamical evolution is found to be restricted by the two first-class constraints
\begin{equation}\label{phi}
\phi_1\equiv p
~~~\mbox{and}~~~
\phi_2 \equiv \alpha\left(xp_y-yp_x\right)+p_z\,.
\end{equation}
Hence, due to the presence of gauge invariance generated by the first-class constraints (\ref{phi}), the quantization of the FLPR model is not so straightforward.  Indeed, depending on the gauge choice, Gribov ambiguities may show up obstructing a direct quantization attempt.  Our goal here is to unravel the BRST-related symmetries and corresponding algebras associated to the FLPR model at quantum level analyzing the role played by Gribov ambiguities in restricting such symmetries along the gauge-fixing process.   To achieve a consistent quantum description,  in the next section, we shall follow the BFV functional quantization method which extends the phase space with proper anti-commuting variables accounting for the necessary degrees of freedom reduction due to the constraint relations (\ref{phi}).

\section{Generating Functional and BRST Symmetry}
According to the well-established Batalin-Fradkin-Vilkovisky (BFV) quantization scheme \cite{Fradkin:1975cq, Batalin:1977pb, Fradkin:1977xi}, corresponding to the two first class constraints (\ref{phi}), we introduce a pair of Grassmann anticommuting ghost functions (${\cal{C}},\bar{\cal{C}}$) along with their respective canonically conjugated momenta ($\bar{\cal{P}},\cal{P}$)
with ghost numbers gh ${\cal{C}}=$ gh ${\cal{P}}=1=-$gh ${\bar{\cal{C}}}=-$gh $\bar{\cal{P}}$, and
write the 
generating functional in the extended phase space as 
\begin{equation}\label{Zpsi}
Z_\Psi =\int {\cal{D}}\varphi \; \exp (iS_{eff}) \,,
\end{equation}
where ${\cal{D}}\varphi$ represents the functional integration measure over all dynamical functions, i.e.,
\begin{equation}
{\cal{D}}\varphi\equiv{\cal{D}}q\,{\cal{D}} p\,{\cal{D}}x\,{\cal{D}}p_x\,{\cal{D}}y\,{\cal{D}}p_y\,{\cal{D}}z\,{\cal{D}}p_z\,{\cal{D}}{{\cal{C}}}\,{\cal{D}}\bar{\cal{P}}\,{\cal{D}}{\bar{\cal{C}}}\,{\cal{D}}{\cal{P}}
\,,
\end{equation}
and $S_{eff}$ stands for the effective action
\begin{equation}
S_{eff}=\int dt\left ({\dot x} p_x + {\dot y} p_y + {\dot z} p_z +{\dot q} p + {\dot{\cal{C}}} \bar{\cal{P}} 
+{\dot{\bar{\cal{C}}}}{ \cal{P}} 
-H_\Psi \right ) \label{seff}\,.
\end{equation}
The extended Hamiltonian $H_\Psi$ appearing in (\ref{seff}) is obtained from (\ref{Hc}) by including the extra ghost functions and gauge-fixing parts in a BRST-closed form as
\begin{equation}\label{Hpsi}
H_\Psi= \frac{1}{2}\left( p_x^2+p_y^2+p_z^2 \right)+V\left(\rho\right)+\left\{\Omega,\Psi\right\} 
\,,
\end{equation}
with $\Omega$ and $\Psi$ representing, respectively, the BRST charge \cite{Becchi:1974xu, Becchi:1974md, Tyutin:1975qk, Becchi:1975nq} and gauge-fixing fermion \cite{Fradkin:1975cq, Batalin:1977pb, Fradkin:1977xi}.  Hence,
the possibility for different gauge-fixing choices enters in the generating functional (\ref{Zpsi}) precisely by means of the fermionic function $\Psi$, with ghost-number $-1$, appearing in (\ref{Hpsi}) through its anti-Poisson-bracket with the BRST charge.

We may further promote the extended phase space dynamic functions to linear operators acting on a Hilbert space satisfying (anti-)commutation relations directly obtained from the corresponding classical Poisson structure.  Thus, the  non-null quantum fundamental relations, with $\hbar=1$, can be written as
\begin{equation}
\begin{aligned}\label{QR}
\left[ q, p \right]_- &= i\,,\\
\left[ x, p_x \right]_- &= \left[ y, p_y \right]_- = \left[ z, p_z \right]_- =i
\,,\\
 \left[ \cal{C}, \bar{\cal{P}} \right]_+ &=      
\left[ {\bar{\cal{C}}}, {{\cal{P}}} \right]_+ =
~ -i \,,
\end{aligned}
\end{equation} 
with the quantum BRST charge operator given by
\begin{equation}
\Omega_b=i \, {\cal{C}}\left[\alpha\left(xp_y-yp_x\right)+p_z\right]+ i\,{{\cal{P}}}p \label{Ob}
\,.
\end{equation}
If $F(x,p_x,y,p_y,z,p_z,q,p,{\cal C},\bar{\cal{P}} ,{\bar{\cal{C}}}, {{\cal{P}}})$ denotes an operator function with a well-defined Grassmann parity $\epsilon_F$, its quantum BRST variation is defined as
\begin{equation}\label{deltaF}
\delta_b F={\left[F,\Omega_b\right]}_{\pm} \equiv F\Omega_b-(-1)^{\epsilon_F}\Omega_bF \,,
\end{equation}
leading in particular to the fundamental BRST variations 
\begin{equation}\label{deltab}
\begin{array}{llll}
 \delta_b x =\alpha y{\cal{C}}, \quad \quad & & \delta_b y= -\alpha x{\cal{C}}, \quad \quad \ \ \delta_b z=-{\cal{C}}, \\
 \delta_b p_x =\alpha p_y{\cal{C}} ,\quad\quad &&\delta_b p_y =-\alpha p_x{\cal {C}}, \quad \ \ \delta_b p_z= 0 ,\\
 \delta_b q= -{\cal{P}},\quad \quad &&\delta_b p=0, \quad \quad \quad \quad\delta _b {\cal{C}}=0, \\
\delta _b {\bar{\cal{C}}}=p \,, \quad \quad &&\delta _b {{\cal{P}}}=0 \,, \quad\quad \quad \quad 
 \delta _b {\bar{\cal{P}}}=\alpha \left(xp_y-yp_x\right)+p_z  \,.  
\end{array}
\end{equation}
The BRST charge $\Omega_b$ has ghost number $+1$ and is nillpotent, by which we mean
\begin{equation}
\Omega_b^2=0\,,
\end{equation}
and, as a direct consequence, the transformations (\ref{deltab}) are off-shell closed.
To proceed further, we need to make a definite choice for $\Psi$ in (\ref{Zpsi}).  That shall be done in the next section.

\section{Gauge-Fixing}
The FLPR model is a first-class system.  The invariance (\ref{GI}) allows for a corresponding freedom, already encoded in $\Psi$, which must be dealt with along the quantization process. 
The BFV standard gauge-fixing procedure  can be worked out from the choice \cite{Henneaux:1985kr, Henneaux:1992ig}
\begin{equation}\label{Psi}
\Psi= {\bar{\cal{P}}}q +{\bar{\cal{C}}}  \chi \,,
\end{equation}
with $\chi$  representing a ghost-independent function to accommodate for a suitable subsidiary condition for the gauge-invariant FLPR model.   Accordingly, different FLPR gauge-fixings, as discussed for instance in \cite{Friedberg:1995ty, Fujikawa:1995fb, Villanueva:1999st}, can be achieved by specific subsidiary conditions present in $\chi$.
As mentioned in the Introduction, in the recent publication \cite{Malik:2023rze}, R. Malik has shown that the quantum version of the first-order FLPR model enjoys a set of off-shell absolutely anticommuting continuous symmetries.  In this section, we show how those symmetries can be obtained from the more general approach developed here, within the framework proposed in reference \cite{Mandal:2023pdk}, as BRST-related symmetries.
To realize the FLPR model nilpotent (anti-)co-BRST symmetries precisely as in \cite{Malik:2023rze}, we choose a bi-parametric linear gauge  of the form
\begin{equation}\label{g1}
\chi=\frac{\beta}{2}p +\omega^2 z 
\end{equation}
with $\beta$ denoting a free dimensionless gauge parameter and $\omega$ providing the appropriate dimension soundness.   A similar gauge-fixing structure has also been used in \cite{Mandal:2023pdk}.
With the choice (\ref{g1}), we may write the BRST invariant Hamiltonian as
\begin{equation}\label{hatH}
\hat{H} = \frac{1}{2}\left( p_x^2+p_y^2+p_z^2 \right)+q\left[\alpha\left(xp_y-yp_x\right)+p_z\right]+V\left( \rho \right)+ {\cal{P}}{\bar{\cal{P}}}+\omega^2{\cal{C}}{\bar{\cal
{C}}}+ p \left( \frac{\beta}{2}p +\omega^2 z \right)\,,
\end{equation}
while the complete extended gauge-fixed action reads
\begin{eqnarray}\label{Sext}
S_{ext}&=&\int dt\left ( p_x{\dot x}+p_y{\dot y}+p_z{\dot z}+ {\dot{{\cal C}}}{{\bar{\cal{P}}}} + {\dot{\bar{\cal C}}}{{{\cal{P}}}} - \frac{1}{2}\left(p_x^2+p_y^2+p_z^2\right) - V(\rho)  \right.  \nonumber \\ 
&& \left. -q\left[\alpha\left(xp_y-yp_x\right)+p_z\right]  - \frac{\beta}{2}p^2+ p{\left( {\dot{q}}-\omega^2 z\right)}
-\omega^2 {\cal C}{\bar{\cal{C}}} - {\cal P}{\bar{\cal{P}}}\right)\,.
\end{eqnarray}
It can be immediately checked that (\ref{Sext}) is invariant under the BRST transformations (\ref{deltab}).
Next, we investigate the discrete group of canonical symmetries produced by the exchange of ghost functions present in (\ref{Sext}) preserving the bracket structure  (\ref{QR}) and ghost number, as suggested in \cite{Mandal:2023pdk}.  By inspection, we see that $S_{ext}$ is left invariant under the action of the group algebra generated by
\begin{equation}\label{Z4}
\mathbb{Z}_4\times\mathbb{Z}_2 = \,\, <a,b \,|\, a^2=b^4=e, ab=ba> \,.
\end{equation}
where $a$ and $b$ denote symmetry generators given by
\begin{equation}\label{can1}
a:~~~{\cal{C}} \longrightarrow \omega^{-1}{\bar{\cal{P}}} \,, ~~~ {\bar{\cal{C}}} \longrightarrow \omega^{-1}{{\cal{P}}} \,, ~~~ {{\cal{P}}} \longrightarrow \omega{\bar{\cal{C}}}
\,, ~~~   {\bar{\cal{P}}}  \longrightarrow
\omega{\cal{C}}
\,
\end{equation}
and
\begin{equation}\label{can2}
b:~~~{\cal{C}} \longrightarrow  - {\bar{\cal{C}}} \,, ~~~ {\bar{\cal{C}}} \longrightarrow  {\cal{C}}   \,, ~~~ {{\cal{P}}} \longrightarrow {\bar{\cal{P}}} 
\,, ~~~  {\bar{\cal{P}}} \longrightarrow
- {{\cal{P}}} 
\,.
\end{equation}
That provides a tool for constructing new symmetries.
In fact, the action of the group algebra generated by (\ref{Z4}) produces the following additional BRST-related symmetries and charges:
\subsection*{Anti-BRST}   
\begin{equation}\label{antideltab}
\begin{array}{llll}
 {\bar\delta_b} x =-\alpha y{\bar{\cal{C}}}, \quad \quad & & {\bar\delta_b} y= \alpha x{\bar{\cal{C}}}, \quad \quad \ \ {\bar\delta_b} z={\bar{\cal{C}}}, \\
 {\bar\delta_b} p_x =-\alpha p_y{\bar{\cal{C}}} ,\quad\quad &&{\bar\delta_b} p_y =\alpha p_x{\bar{\cal {C}}}, \quad \ \ {\bar\delta_b} p_z= 0 ,\\
 {\bar\delta_b} q= -{\bar{\cal{P}}},\quad \quad &&{\bar\delta_b} p=0, \quad \quad \quad \quad{\bar\delta _b} {\cal{C}}=p, \\
{\bar\delta _b} {\bar{\cal{C}}}=0 \,, \quad \quad &&{\bar\delta _b} {\bar{\cal{P}}}=0 \,, \quad\quad \quad \quad 
 {\bar\delta _b} {\cal{P}}=\alpha\left(xp_y-yp_x\right)+p_z \,. 
\end{array}
\end{equation}
\begin{equation}\label{antiOmegab}
\bar{\Omega}_b=-i{\bar{\cal{C}}}\left[\alpha \left(xp_y-yp_x\right)+p_z\right] + i{\bar{{\cal{P}}}}p \,, ~~~~ \mbox{ gh }\bar{\Omega}_b=-1\,.
\end{equation}
\subsection*{Dual-BRST} 
\begin{equation}\label{deltad}
\begin{array}{llll}
 {\bar\delta_d} x =\alpha \omega^{-1}y{\bar{\cal{P}}}, \quad \quad & & {\bar\delta_d} y= -\alpha\omega^{-1}x{\bar{\cal{P}}}, \quad \quad \ \ {\bar\delta_d} z=-\omega^{-1}{\bar{\cal{P}}}, \\
 {\bar\delta_d} p_x =\alpha\omega^{-1}p_y{\bar{\cal{P}}} ,\quad\quad &&{\bar\delta_d} p_y =-\alpha \omega^{-1}p_x{\bar{\cal {P}}}, \quad \ \ {\bar\delta_d} p_z= 0 ,\\
 {\bar\delta_d} q= -\omega{\bar{\cal{C}}},\quad \quad &&{\bar\delta_d} p=0, \quad \quad \quad \quad\quad \quad {\bar\delta _d} {\cal{P}}=\omega p, \\
{\bar\delta _d} {\bar{\cal{C}}}=0 \,, \quad \quad &&{\bar\delta _d} {\bar{\cal{P}}}=0 \,, \quad\quad \quad \quad \quad\quad
 {\bar\delta _d} {\cal{C}}=\omega^{-1}\left[\alpha \left(xp_y-yp_x\right)+p_z\right]\,,
\end{array}
\end{equation}
\begin{equation}\label{Omegad}
\bar{\Omega}_d=-i\omega^{-1}{\bar{\cal{P}}}\left[\alpha\left(xp_y-yp_x\right)+p_z\right] + i\omega{\bar{{\cal{C}}}}p\,, ~~~~ \mbox{ gh }\bar{\Omega}_d=-1\,.
\end{equation}
\subsection*{Anti-dual BRST}  
\begin{equation}\label{antideltad}
\begin{array}{llll}
 \delta_d x =-\alpha\omega^{-1}y{\cal{P}}, \quad \quad & & \delta_d y= \alpha\omega^{-1}x{\cal{P}}, \quad \quad \ \ \delta_d z=\omega^{-1}{\cal{P}}, \\
 \delta_d p_x =-\alpha\omega^{-1}p_y{\cal{P}} ,\quad\quad &&\delta_d p_y =\alpha\omega^{-1}p_x{\cal {P}}, \quad \ \ \delta_d p_z= 0 ,\\
 \delta_d q= -\omega{\cal{C}},\quad \quad &&\delta_d p=0, \quad \quad \quad \quad\quad  \delta _d {\bar{\cal{P}}}=\omega p, \\
 \delta _d {\cal{C}}=0 \,, \quad \quad &&\delta _d {\cal{P}}=0 \,, \quad\quad \quad \quad \quad
 \delta _d {\bar{\cal{C}}}=-\omega^{-1}\left[\alpha\left(xp_y-yp_x\right)+p_z\right]\,,
\end{array}
\end{equation}
\begin{equation}\label{antiOmegad}
{\Omega}_d=-i\omega^{-1}{{\cal{P}}}\left[\alpha\left(xp_y-yp_x\right)+p_z\right]+i\omega {{\cal{C}}}p
\,, ~~~~ \mbox{ gh }{\Omega}_d=+1\,.
\end{equation}
The above BRST-related symmetries (\ref{antideltab}), (\ref{deltad}) and (\ref{antideltad}) are all off-shell closed, each one respectively possessing a conserved nill-potent fermionic charge  (\ref{antiOmegab}), (\ref{Omegad}) and (\ref{antiOmegad}).   To unravel and perfectly match the symmetries identified in \cite{Malik:2023rze},  we functionally integrate  (\ref{Zpsi}) in the ghost momenta to obtain the first-order action
\begin{eqnarray}\label{Sfo}
S_{fo}&=&\int dt\left ( p_x{\dot x}+p_y{\dot y}+p_z{\dot z}- \frac{1}{2}\left(p_x^2+p_y^2+p_z^2\right) - V(\rho)  \right.  \nonumber \\ 
&& \left. -q\left[\alpha\left(xp_y-yp_x\right)+p_z\right]  - \frac{\beta}{2}p^2+ p{\left( {\dot{q}}-\omega^2 z\right)}
+{\dot{\bar{\cal C}}}{\dot{{\cal{C}}}}
-\omega^2 {\cal C}{\bar{\cal{C}}} 
\right)\,.
\end{eqnarray}
As a consequence, the previous symmetries become now
\subsection*{BRST}
\begin{equation}\label{sb}
\begin{array}{lll}
 s_b x =\alpha y{\cal{C}}, \quad \quad & & s_b y= - \alpha x{\cal{C}}, \quad \quad \ \ s_b z=-{\cal{C}}, \\
 s_b p_x =\alpha p_y{\cal{C}} ,\quad\quad &&s_b p_y =- \alpha p_x{\cal {C}}, \quad \ \ s_b p_z= 0 ,\\
 s_b q= -{\dot{\cal{C}}},\quad \quad &&s_b p=0, \quad \quad \quad \quad s _b {\cal{C}}=0, \\
s_b {\bar{\cal{C}}}=p \,, 
\end{array}
\end{equation}
\begin{equation}\label{Qb}
{\cal Q}_b=i{\cal{C}}\left[\alpha\left(xp_y-yp_x\right)+p_z\right] + i{\dot{\cal{C}}}p ~~~~ \mbox{ gh }{{\cal Q}}_b=+1\,,
\end{equation}
\subsection*{Anti-BRST}
\begin{equation}\label{antisb}
\begin{array}{lll}
 {\bar{s}}_b x =-\alpha y{\bar{\cal{C}}}, \quad \quad & & {\bar{s}}_b y= \alpha x{\bar{\cal{C}}}, \quad \quad \ \ {\bar{s}}_b z={\bar{\cal{C}}}, \\
 {\bar{s}}_b p_x =-\alpha p_y{\bar{\cal{C}}} ,\quad\quad && {\bar{s}}_b p_y =\alpha p_x{\bar{\cal {C}}}, \quad \ \ {\bar{s}}_b p_z= 0 ,\\
 {\bar{s}}_b q= {\dot{\bar{\cal{C}}}},\quad \quad &&{\bar{s}}_b p=0, \quad \quad \quad \quad {\bar{s}} _b {\cal{C}}=p, \\
s_b {\bar{\cal{C}}}=0  \,,
\end{array}
\end{equation}
\begin{equation}\label{antiQb}
{\bar{\cal Q}}_b=-i{\bar{\cal{C}}}\left[g\left(xp_y-yp_x\right)+p_z\right] + i{\dot{\bar{\cal{C}}}}p \,,
~~~~ \mbox{ gh }\bar{{\cal Q}}_b=-1\,,
\end{equation}
\subsection*{Dual-BRST}
\begin{equation}\label{sd}
\begin{array}{lll}
 {\bar{s}}_d x =-\alpha\omega^{-1}y{\dot{\bar{\cal{C}}}}, \quad \quad & & {\bar{s}}_d y= \alpha\omega^{-1}x{\dot{\bar{\cal{C}}}}, \quad \quad \ \ {\bar{s}}_d z=\omega^{-1}{\dot{\bar{\cal{C}}}}, \\
 {\bar{s}}_d p_x =-\alpha\omega^{-1}p_y{\dot{\bar{\cal{C}}}} ,\quad\quad && {\bar{s}}_d p_y =\alpha\omega^{-1}p_x{\dot{\bar{\cal {C}}}}, \quad \ \ {\bar{s}}_d p_z= 0 ,\\
 {\bar{s}}_d q=-\omega {\bar{\cal{C}}},\quad \quad &&{\bar{s}}_d p=0, \quad \quad \quad \quad {\bar{s}} _d {\cal{C}}=\omega^{-1}\left[\alpha \left(xp_y-yp_x \right)+p_z \right], \\
{\bar{s}}_d {\bar{\cal{C}}}=0 
\end{array}
\end{equation}
\begin{equation}\label{Qd}
{\bar{\cal Q}}_d=-i\omega^{-1}{\dot{\bar{\cal{C}}}}\left[\alpha \left(xp_y-yp_x\right)+p_z\right] + i\omega{\bar{\cal{C}}}p 
\,, ~~~~ \mbox{ gh }\bar{{\cal Q}}_d=-1\,,
\end{equation}
\subsection*{Anti-Dual-BRST}
\begin{equation}\label{antisd}
\begin{array}{lll}
 {s}_d x =-\alpha\omega^{-1}y{\dot{\cal{C}}}, \quad \quad & & {s}_d y= \alpha\omega^{-1}x{\dot{\cal{C}}}, \quad \quad \ \ {s}_d z=\omega^{-1}{\dot{\cal{C}}}, \\
 {s}_d p_x =-\alpha\omega^{-1}p_y{\dot{\cal{C}}} ,\quad\quad && {s}_d p_y =\alpha\omega^{-1}p_x{\dot{\cal {C}}}, \quad \ \ {s}_d p_z= 0 ,\\
 {s}_d q=\omega {\cal{C}},\quad \quad &&{s}_d p=0, \quad \quad \quad \quad {s} _d {\bar{\cal{C}}}=-\omega^{-1}\left[\alpha\left(xp_y-yp_x \right)+p_z \right], \\
{s}_d {\cal{C}}=0 
\end{array}
\end{equation}
\begin{equation}\label{antiQd}
{{\cal Q}}_d=-i\omega^{-1}{\dot{\cal{C}}}\left[\alpha\left(xp_y-yp_x\right)+p_z\right] + i\omega{\cal{C}}p 
\,, ~~~~ \mbox{ gh }{{\cal Q}}_d=+1\,.
\end{equation}
All the above transformations are off-shell nillpotent and each one of them can be checked to characterize a particular symmetry of the action (\ref{Sfo}).  Furthermore, the two dual-BRST symmetries (\ref{sd}) and (\ref{antisd}) do leave the gauge-fixing term invariant, as expected from a genuine {\it dual-} or {\it co-}BRST transformation \cite{Malik:1997ge, Mandal:2023pdk}. 

\section{Gribov Ambiguities}
Naturally, there are other possibilities for gauge-fixing the degrees of freedom of the FLPR model.   The main motivation and purpose of the FLPR model, as a toy model mimicking certain specific tricky aspects of QCD, lies on its intrinsic relation to the Gribov problem \cite{Gribov:1977wm}. In references \cite{Friedberg:1995ty, Fujikawa:1995fb}, it has been shown that the subsidiary condition
\begin{equation}\label{Ggauge}
z-\lambda x = 0
\end{equation}
is plagued by Gribov ambiguities in a seemingly analogous form to the Landau gauge in QCD.  In case (\ref{Ggauge}) holds, then for $\lambda\alpha y = -1$, any infinitesimal transformation (\ref{GI}) leads to another configuration still satisfying (\ref{Ggauge}).  For finite transformations, again assuming (\ref{Ggauge}), all solutions of
\begin{equation}
x[1-\cos(\alpha\xi)]+y\sin(\alpha\xi)=-\xi/\lambda
\end{equation}
lead to the same gauge condition. 
In order to implement (\ref{Ggauge}) in our formalism, we choose
$\Psi$ in (\ref{Psi}) as
\begin{equation}\label{g2}
\chi=\frac{\beta}{2}p +\omega^2 (z-\lambda x) +\dot{q}\,. 
\end{equation}
and write the corresponding gauge-fixed action as
\begin{eqnarray}\label{SG}
S_{G}&=&\int dt\left ( p_x{\dot x}+p_y{\dot y}+p_z{\dot z} + {\dot{{\cal C}}}{{{\bar{\cal{P}}}}} - \frac{1}{2}\left(p_x^2+p_y^2+p_z^2\right) - V(\rho)  \right.  \nonumber \\ 
&& \left. -q\left[\alpha\left(xp_y-yp_x\right)+p_z\right]  - \frac{\beta}{2}p^2- \omega^2 p{\left( z - \lambda x \right)} 
- {\cal P}{\bar{\cal{P}}} -\omega^2 {\cal C}(1+\alpha \lambda y){\bar{\cal{C}}} \right) \,.
\end{eqnarray}
Note that the last term in the integrand of (\ref{SG}) plays a similar role to its analogous in QCD, containing a field-dependent Faddeev-Popov determinant whose first zero-mode gives the horizon of the Gribov region \cite{Capri:2006cz}, inside of which there are no Gribov copies.
Although action (\ref{SG}) is BRST invariant under ({\ref{deltab}}), it is not invariant under the BRST-related transformations (\ref{antideltab}), (\ref{deltad}) or (\ref{antideltad}) anymore, abiding to the fact that the discrete symmetry (\ref{Z4}) is now broken.    

In lieau of (\ref{can1}) and (\ref{can2}), it can be seen by inspection that 
\begin{equation}\label{can3}
c:~~~{\cal{C}} \longrightarrow \omega^{-1}{\bar{\cal{P}}} \,, ~~~ {\bar{\cal{C}}} \longrightarrow -\omega^{-1}(1+\alpha\lambda y)^{-1}{{\cal{P}}} \,, ~~~ {{\cal{P}}} \longrightarrow -\omega(1+\alpha\lambda y){\bar{\cal{C}}}
\,, ~~~   {\bar{\cal{P}}}  \longrightarrow
\omega{\cal{C}}
\,
\end{equation}
constitutes an invertible discrete symmetry of (\ref{SG}) leading to the BRST-related transformation 
\begin{equation}\label{Ndeltad}
\begin{array}{llll}
 {\bar\delta_d}' x =\alpha \omega^{-1}y{\bar{\cal{P}}}, \quad \quad & & {\bar\delta_d}' y= -\alpha\omega^{-1}x{\bar{\cal{P}}}, \quad \quad \ \ {\bar\delta_d}' z=-\omega^{-1}{\bar{\cal{P}}}, \\
 {\bar\delta_d}' p_x =\alpha\omega^{-1}p_y{\bar{\cal{P}}} ,\quad\quad &&{\bar\delta_d}' p_y =-\alpha \omega^{-1}p_x{\bar{\cal {P}}}, \quad \ \ {\bar\delta_d}' p_z= 0 ,\\
 {\bar\delta_d}' q= \omega(1+\alpha\lambda y){\bar{\cal{C}}},\quad \quad &&{\bar\delta_d}' p=0, \quad \quad \quad \quad\quad \quad {\bar\delta _d}' {\cal{P}}=-\omega(1+\alpha\lambda y) p+\displaystyle\frac{\alpha^2\omega^{-1} \lambda x {\bar{\cal{P}}}{\cal P}}{1+\alpha \lambda y} \,, \\
{\bar\delta _d}' {\bar{\cal{C}}}=-\displaystyle\frac{\alpha^2\omega^{-1}\lambda x}{1+\alpha\lambda y}{\bar{\cal{P}}}{\bar{\cal{C}}} \,, \quad \quad &&{\bar\delta _d}' {\bar{\cal{P}}}=0 \,, \quad\quad \quad \quad \quad\quad
 {\bar\delta _d}' {\cal{C}}=\omega^{-1}\left[\alpha \left(xp_y-yp_x\right)+p_z\right]\,.
\end{array}
\end{equation}
By construction, as generated by (\ref{can3}), the ${\bar\delta_d}'$ transformation (\ref{Ndeltad}) characterizes a BRST-related symmetry of the action (\ref{SG})
\begin{equation}
{\bar\delta_d}' S_G = 0
\,,
\end{equation}
as can be explicitly checked out.

The Gribov-fixed action (\ref{SG}) furnishes a starting point for a functional quantization of the model whose restriction to the Gribov region defined by
\begin{equation}\label{GR}
1+\alpha\lambda y >0
\end{equation}
is well-defined.  On those grounds, using (\ref{SG}), we may integrate (\ref{Zpsi}) in the ghost momenta to obtain a Gribov-fixed effective first-order action
\begin{eqnarray}\label{SGfo}
S_{Gfo}&=&\int dt\left ( p_x{\dot x}+p_y{\dot y}+p_z{\dot z}  - \frac{1}{2}\left(p_x^2+p_y^2+p_z^2\right) - V(\rho)  \right.  \nonumber \\ 
&& \left. -q\left[\alpha\left(xp_y-yp_x\right)+p_z\right]  - \frac{\beta}{2}p^2- \omega^2 p{\left( z - \lambda x \right)} 
 -\omega^2 {\cal C}(1+\alpha \lambda y){\bar{\cal{C}}} \right)
\end{eqnarray}
which turns out to be invariant under (\ref{sb}) and the integrated version of \ref{Ndeltad}, namely,
\begin{equation}
\begin{array}{lll}
 {\bar{s}}_d' q=&& \omega(1+\alpha \lambda y)\bar{\cal C}\,,
\\
 {\bar{s}}_d' {\cal C}=&& \omega^{-1}\left[ \alpha\left(xp_y-yp_x\right)+p_z \right]
\,.
\end{array}
\end{equation}
In this way, the modified partition function avoiding Gribov copies can be written as
\begin{equation}
{\cal Z} =\int {\mathbf{\tilde{\cal{D}}}}\varphi \; \exp (iS_{Gfo}) \,,
\end{equation} 
with
\begin{equation}
{\mathbf{\tilde{\cal{{D}}}}} 
\varphi\equiv {\mathcal{V}}(C_0) \, {\cal{D}}q\,{\cal{D}} p\,{\cal{D}}x\,{\cal{D}}p_x\,{\cal{D}}y\,{\cal{D}}p_y\,{\cal{D}}z\,{\cal{D}}p_z\,{\cal{D}}{{\cal{C}}}\,{\cal{D}}{\bar{\cal{C}}}\,,
\end{equation}
where the term ${\mathcal{V}}(C_0)$ in the measure signals that the functional integration is restricted to the space phase Gribov region characterized by (\ref{GR}). 
In this way, we have succeeded in obtaining a prescription for a consistent quantization of the FLPR model in a Gribov gauge by avoiding the Gribov copies with corresponding BRST-related symmetries.

\section{Conclusion}
We have shown that a class of BRST-related symmetries can be systematically generated in the FLPR model.  The set of nillpotent continuous symmetries and corresponding algebra presented in \cite{Malik:2023rze} has been shown to be reproducible by the group algebra generated by the discrete symmetries of the gauge-fixed action without Gribov ambiguities.  However, we have seen that the asymmetry present in the Gribov gauge-fix choice considerably reduced the group algebra of discrete symmetries of the action, consequently breaking the associated BRST-related symmetries.  In that case, we have been able to partially restore the BRST-related symmetries in terms of the ${\bar\delta_d}'$ transformation.  This corresponds to a similar situation as in QCD quadratic gauges, in which the Faddeev-Popov action does not possess anti-BRST symmetry \cite{Joglekar:1994tq}.  The functional quantization can be done avoiding Gribov copies, by restricting the integration space to the analogous of the Gribov region in the FPLR model.

{\bf Acknowledgements:} B.P.M. acknowledges the incentive research grant for faculty under the IoE Scheme (IoE/Incentive/2021-22/32253) of the Banaras Hindu University.  S.K.R. thanks the Department of Biotechnology, Government of India, New Delhi for partial financial support.

\end{document}